\documentclass[graybox]{svmult}
\usepackage[utf8]{inputenc}
\usepackage[T1]{fontenc}
\usepackage[finnish, english]{babel}
\usepackage{mathptmx}       % selects Times Roman as basic font
\usepackage{helvet}         % selects Helvetica as sans-serif font
\usepackage{courier}        % selects Courier as typewriter font
\usepackage{makeidx}         % allows index generation
\usepackage{graphicx}        % standard LaTeX graphics tool
\graphicspath{ {images/} }
\usepackage{multicol}        % used for the two-column index
\usepackage[bottom]{footmisc}% places footnotes at page bottom
\usepackage{amsmath,amssymb,amsfonts}        % AMS math packages

\usepackage{url}   

\newcommand{\R}{\mathbb{R}}
\newcommand{\X}{\mathbb{X}}

\newcommand{\II}{\mathbb{I}}

\newcommand{\Ex}{\mathbb{E}}

\newcommand{\N}{\mathcal{N}}
\newcommand{\Spr}{\Sigma_{pr}}
\newcommand{\Snoise}{\Sigma_{obs}}

\newcommand{\Spri}{\Sigma_{pr}^{-1}}
\newcommand{\Snoisei}{\Sigma_{obs}^{-1}}

\newcommand{\Lc}{\mathcal{L}}

\makeindex             % used for the subject index
\begin{document}
%\title*{Dimension reduction in atmospheric remote sensing}
\title*{Likelihood informed dimension reduction for inverse problems in remote sensing of atmospheric constituent profiles}
\titlerunning{LIS dimension reduction for atmospheric remote sensing}
\author{Otto Lamminpää, Marko Laine, Simo Tukiainen, Johanna Tamminen}
\institute{Otto Lamminpää 
\and Marko Laine 
\and Simo Tukiainen 
\and Johanna Tamminen \at Finnish Meteorological Institute, P.O. BOX 503,
FI-00101 Helsinki, Finland.  \email{otto.lamminpaa@fmi.fi}}

% Keywords: Likelihood informed dimension reduction; statistical inverse problems; remote sensing; atmospheric profiles; methane; adaptive MCMC

\maketitle

\abstract{
%\newline\indent
We use likelihood informed dimension reduction (LIS) \cite{TC1} for inverting vertical profile information of atmospheric methane from ground based Fourier transform infrared (FTIR) measurements at Sodankylä, Northern Finland. The measurements belong to the word wide TCCON network for greenhouse gas measurements and, in addition to providing accurate greenhouse gas measurements, they are important for validating satellite observations.
\newline\indent
LIS allows construction of an efficient Markov chain Monte Carlo sampling algorithm that explores only a reduced dimensional space but still produces a good approximation of the original full dimensional Bayesian posterior distribution. This in effect makes the statistical estimation problem independent of the discretization of the inverse problem.
In addition, we compare LIS to a dimension reduction method based on prior covariance matrix truncation used earlier \cite{Simo}.
}

\section{Introduction}
\label{sec:1}
Atmospheric composition measurements have an increasingly crucial role in monitoring the green house gas concentrations in order to understand and predict changes in climate. The warming effect of greenhouse gases, such as carbon dioxide ($\text{CO}_2$) and methane ($\text{CH}_4$), is based on the absorption of electromagnetic radiation originating from the sun by these trace gases. This mechanism has a strong theoretical base and has been confirmed by recent observations \cite{FeldmanEtal15}.

Remote sensing measurements of atmospheric composition, and greenhouse gases in particular, are carried out by ground-based Fourier transform infrared (FTIR) spectrometers, and more recently by a growing number of satellites (for example SCIAMACHY, ACE-FTS, GOSAT, OCO-2). The advantage of satellite measurements is that they provide global coverage. They are used for anthropogenic emission monitoring, detecting trends in atmospheric composition and studying the effects of biosphere, to name but a few examples. Accurate ground-based measurements are crucial to satellite measurement validation, and the global Total Carbon Column Observing Network (TCCON \cite{WunchEtAl15}) of FTIR spectrometers, consisting of around 20 measurement sites around the world, is widely used as a reference \cite{DilsEtAl14}. The FTIR instrument looks directly at sun, returning an absorption spectrum as measured data.

Determining atmospheric gas density profiles, or \emph{retrieval}, from the absorption spectra is an ill-defined \emph{inverse problem} as the measurement contains only a limited amount of information about the state of the atmosphere. Based on prior knowledge and using the Bayesian approach to regularize the problem, the profile retrieval is possible, provided that our prior accurately describes the possible states that may occur in the atmosphere. When retrieving a vertical atmospheric profile, the dimension of the estimation problem depends on the discretization. For accurate retrievals a high number of layers are needed, leading to a computationally costly algorithms. However, fast methods are required for the operational algorithm. For this purpose, different ways of reducing the dimension of the problem have been developed. The official operational TCCON GGG algorithm \cite{WunchEtAl15} solves the inverse problem by scaling the prior profile based on the measured data. This method is robust and computationally efficient, but only retrieves one piece of information and thus can give largely inaccurate results about the density profiles.

An improved dimension reduction method for the FTIR retrieval based on reducing the rank of the prior covariance matrix was used by Tukiainen et al. \cite{Simo} using computational methods developed by Solonen et al. \cite{Solonen}. This method confines the solution to a subspace spanned by the non-negligible eigenvectors of the prior covariance matrix. This approach allows a retrieval using more basis functions than the operational method and thus gives more accurate solutions. However, the prior has to be hand tuned to have a number of non-zero singular values that correspond to the number of degrees of freedom for the signal in the measurement. Moreover, whatever information lies in the complement of this subspace remains unused. 

In this work, we introduce an analysis method for determining the number of components the measurement can provide information from \cite{Rodgers}, as well as the \emph{likelihood informed subspace} dimension reduction method for non-linear statistical inverse problems \cite{TC1,TC2}. We show that these two formulations are in fact equal. We then proceed to implement a dimension reduction scheme for the FTIR inverse problem using adaptive MCMC sampling \cite{HaarioEtAl1,HaarioEtAl2} to fully characterize the non-linear posterior distribution, and show that this method gives an optimal result with respect to Hellinger distance to the non-approximated full dimensional posterior distribution. In contrast with the previously implemented prior reduction method, the likelihood informed subspace method is also shown to give the user freedom to use a prior derived directly from an ensemble of previously conducted atmospheric composition measurements. 

\section{Methodology}
\label{sec:2}
We consider the atmospheric composition measurement carried out at the FMI Arctic Research Centre, Sodankylä, Finland \cite{KiviEtAl16}. The on-site Fourier transform infrared spectrometer (FTIR) measures solar light arriving to the device directly from the sun, or more precisely, the absorption of solar light at different wavelengths within the atmosphere. From the absorption spectra of different trace gases ($\text{CO}_2, \text{CH}_4$, $\ldots$) we can compute the corresponding vertical density profiles, i.e.\ the fraction of the trace gas in question as a function of height.

Let us consider the absorption spectrum with $m$ separate wavelengths. The solar light passing through the atmosphere and hitting the detector can be modeled using the \emph{Beer-Lambert law}, which gives, for wavelengths $\lambda_j, j\in[1, \dots, m]$, the intensity of detected light as 
\begin{equation}\label{lambeer}
	I(\lambda_j) = I_0(\lambda_j)\exp\left( -\sum_{k=1}^K\int_0^{\infty}\mathcal{C}_k(\lambda_j,z)\rho_k(z)dz \right)(a\lambda_j^2+b\lambda_j+c) + d,
\end{equation}
where $I_0$ is the intensity of solar light when it enters the atmosphere, the atmosphere has $K$ absorbing trace gases,  $\mathcal{C}_k(\lambda_j, z)$ is the absorption coefficient of gas $k$, which depends on height $z$ and on the wavelength $\lambda_j$, and $\rho_k(z)$ is the density of gas $k$ at height $z$. The second degree polynomial and the constant $d$ in (\ref{lambeer}) are used to describe instrument related features and the continuity properties of the spectrum. 
In reality, solar light is scattered on the way by atmospheric particles. This phenomenon is relatively weak in the wavelength band we are considering in this work, so it will be ignored for simplicity.

The absorption in continuous atmosphere is modeled by discretizing the integral in equation (\ref{lambeer}) into a sum over atmospheric layers and assuming a constant absorption for each separate layer. This way, a discrete computational \emph{forward model} can be constructed, giving an absorption spectrum as data produced by applying the forward model to a state vector $x$ describing the discretized atmospheric density profile for a certain trace gas. In this work, we limit ourselves to consider the retrieval of atmospheric methane ($\text{CH}_4$).

\subsection{Bayesian formulation of the inverse problem}
Consider an inverse problem of estimating unknown parameter vector $x\in\R^n$ from observation $y\in\R^m$,
\begin{equation}\label{inverseproblem}
	y=F(x)+\varepsilon,
\end{equation}
where our physical model is describe by the \emph{forward model} $F: \R^n\rightarrow\R^m$ and the random variable $\varepsilon\in\R^m$ represents the observation error arising from instrument noise and forward model approximations. In the Bayesian approach to inverse problems \cite{KaipioSomersalo} our uncertainty about $x$ is described by statistical distributions. The solution to the problem is obtained as posterior distribution of $x$ conditioned on a realization of the data $y$ and depending on our prior knowledge. By the Bayes' formula, we have
\begin{equation}\label{bayes}
  \pi(x|y) \propto \pi(y|x)\pi_{pr}(x),
\end{equation}
where $\pi(x|y)$ is the posterior distribution, $\pi(y|x)$ the likelihood and $\pi_{pr}(x)$ the prior distribution. The proportionality $\propto$ comes from a constant that does not depend on the unknown $x$. In this work, we assume the prior to be Gaussian, $\N(x_0,\Spr)$, e.g. 
\begin{equation}
  \pi_{pr}(x) \propto \exp\left(-\frac{1}{2}(x-x_0)^T\Spri(x-x_0)\right).
\end{equation}
Also, the additive noise is assumed to be zero-mean Gaussian with known covariance matrix,  $\varepsilon \sim \N(0, \Snoise)$, so the likelihood will have form
\begin{equation}
  \pi(y|x) \propto \exp\left(-\frac{1}{2}(y-F(x))^T\Snoisei(y-F(x))\right).
\end{equation}
When the forward model is non-linear, the posterior distribution can be explored by Markov chain Monte Carlo (MCMC) sampling. When the dimension of the unknown is hight, for example by discretization of the inverse problem, MCMC is known to be inefficient. In this paper, we utilize dimension reduction to be able to make MCMC more efficient in high dimensional and high CPU problems.

\subsection{Prior reduction}
The operational GGG algorithm for the FTIR retrieval problem \cite{WunchEtAl15} is effectively one dimensional as it only scales the prior mean profile. However, there are about three degrees of freedom in the FTIR signal for the vertical profile information. To construct basis functions that could utilize this information a method that uses prior reduction was developed in \cite{Simo}.
It is based on the singular value decomposition on the prior covariance matrix,
\begin{equation}
	\Spr = U\Lambda U^T = \sum_{i=1}^m \lambda_i u_i u_i^T,
\end{equation}
which allows further decomposition as
\begin{equation}
\Spr = PP^T, \text{ with } P = \left( \sqrt{\lambda_1}u_1 + \cdots + \sqrt{\lambda_m}u_m\right).
\end{equation}
If the prior can be chosen so that most of the singular values are negligible, then the rank of the prior covariance matrix can be reduced by considering only the first $r$ singular values and vectors:
\begin{equation}
	\widetilde{\Sigma}_{pr} = P_rP^T_r, \text{ with } P_r = \left( \sqrt{\lambda_1}u_1 + \cdots + \sqrt{\lambda_r}u_r\right).
\end{equation}
The unknown $x$ has an approximate representation by $r$ basis vectors from the columns of $P_r$ and using a reduced dimensional parameter $\alpha \in \R^r$ as 
\begin{equation}\label{eq:alpha}
  x \approx x_0 + P_r\alpha.
\end{equation}
By the construction, the random vector $\alpha$ has a simple Gaussian prior, $\alpha \sim \N(0, \II)$, which allow us to write the approximate posterior as
\begin{equation}
	\pi(x|y) \approx \widetilde{\pi}(\alpha|y) \propto \exp\left( -\frac{1}{2}\left((y -F(x_0+P_r\alpha))^T\Snoisei(y -F(x_0+P_r\alpha)) + \alpha^T\alpha) \right) \right).
\end{equation}
Now, instead running MCMC in the full space defined by $x$, we can sample the low dimensional parameter $\alpha$ and retain the approximation of the full posterior by equation~(\ref{eq:alpha}). 

\subsection{Likelihood-informed subspace}

The prior reduction approach depends on the ability to construct a realistic prior that can be described by only a few principle components. For the FTIR retrieval problem this is possible to some extent \cite{Simo}. However, there are several possible caveats. We have to manually manipulate the prior covariance matrix to have a lower rank, which can lead to information loss as the solution will be limited to a subspace defined by the reduced prior only. 

In atmospheric remote sensing the information content of the measurement is an important concept to be considered when designing the instruments and  constructing the retrieval methodology, we refer to book by Rodgers \cite{Rodgers}.

Consider a linearized version of the inverse problem in equation (\ref{inverseproblem}),
\begin{equation}
  y = J(x-x_0) + \varepsilon,
\end{equation}
with Gaussian prior and noise. The forward model is assumed to be differentiable, and $J$ denotes the Jacobian matrix of the forward model with elements $J_{ij} = \frac{\partial}{\partial x_j}F_i$. Using Cholesky factorizations for the known prior and error covariances,
\begin{equation}
  \Spr = \Lc_{pr}\Lc_{pr}^T, \quad\Snoise = \Lc_{obs}\Lc_{obs}^T,
\end{equation}
we can perform pre-whitening of the problem by setting 
\begin{equation}
\widetilde{y} = \Lc_{obs}^{-1}y,\quad
  \widetilde{J} = \Lc_{obs}^{-1}J\Lc_{pr}, \quad \widetilde{x} = \Lc_{pr}^{-1}(x-x_0) \text{ and } \widetilde{\varepsilon} = \Lc_{obs}^{-1}\varepsilon.
\end{equation}
Now the problem can be written as
\begin{equation}
  \widetilde{y} = \widetilde{J}\widetilde{x}+\widetilde{\varepsilon},
\end{equation}
with $\widetilde{\varepsilon}\sim \N(0,\II)$ and a priori $\widetilde{x}\sim \N(0,\II)$.

As the unknown $x$ and the error $\varepsilon$ are assumed to be independent, the same holds for the scaled versions. We can compare the prior variability of the observation depending on $x$ and that coming from the noise $\varepsilon$ by
\begin{equation}
  \widetilde{\Sigma}_{y}=\Ex[\widetilde{y}\widetilde{y}^T] = \Ex[(\widetilde{J}\widetilde{x}+\widetilde{\varepsilon})(\widetilde{J}\widetilde{x}+\widetilde{\varepsilon})^T]=\widetilde{J}\widetilde{J}^T+\II.
\end{equation}
The variability in $y$ that depends only on the parameter $x$ depends itself on $\widetilde{J}\widetilde{J}^T$ and it can be compared to the unit matrix $\II$ that has the contribution from the scaled noise. The directions in $\widetilde{J}\widetilde{J}^T$ which are larger than unity are those dominated by the signal. Formally this can be seen by diagonalizing the scaled problem by the singular value decomposition,
\begin{equation}\label{svd}
	\widetilde{J} = W\Lambda V^T,
\end{equation} 
and setting
\begin{equation}
	y^{\prime} = W^T\widetilde{y}=W^T\widetilde{J}\widetilde{x} + W^T\widetilde{\varepsilon} = \Lambda V^T\widetilde{x}+\widetilde{\varepsilon}^{\prime} = \Lambda\widetilde{x}^{\prime}+\widetilde{\varepsilon}^{\prime}.
\end{equation}
The transformations $\varepsilon^{\prime}$ and $x^{\prime}$ conserve the unit covariance matrix. In other words, $y^{\prime}$ is distributed with covariance $\Lambda^2+\II$. This is a diagonal matrix, and the elements of vector $y^{\prime}$ that are not masked by the measurement error are those corresponding to the singular values $\lambda_i\geq 1$ of the pre-whitened Jacobian $\widetilde{J}$. Furthermore, degrees of freedom for signal and noise are invariant under linear transformations \cite{Rodgers}, so the same result is also valid for the original $y$.

Another way to compare the information content of the measurement relative to the prior was used in \cite{TC1}. This is to use the Rayleigh quotient
\begin{equation}\label{raylei}
	\mathcal{R}(\Lc_{pr}a) = \frac{a^T\Lc_{pr}^TH\Lc_{pr}a}{a^Ta},
\end{equation}
where $a \in \R^n$ and $H=J^T\Snoise^{-1}J$ is the Gauss-Newton approximation of Hessian matrix of the data misfit function
\begin{equation}
  \label{equ:eta}
\eta(x) = \frac{1}{2}\left((y-F(x))^T\Snoisei(y-F(x))\right).
\end{equation}
Directions for which $\mathcal{R}(\Lc_{pr}a)>1$ are the ones in which the likelihood contains information relative to the prior. This follows from the fact that the $i$th eigenvector $v_i$ of the prior-preconditioned Gauss-Newton Hessian 
\begin{equation}
  \label{equ:hess}
\widetilde{H} := \Lc_{pr}^TH\Lc_{pr}  
\end{equation}
maximizes the Rayleigh quotient over a subspace $\R^n \setminus \text{span}\left\{v_1, \dots, v_{i-1} \right\}$ and the $r$ directions $v_i$ for which $\mathcal{R}(\Lc_{pr}v)>1$ correspond to the first $r$ eigenvalues of $\widetilde{H}$. We call these vectors the \emph{informative directions of the measurement}.

\begin{figure}
\centering
\includegraphics[width=0.8\textwidth]{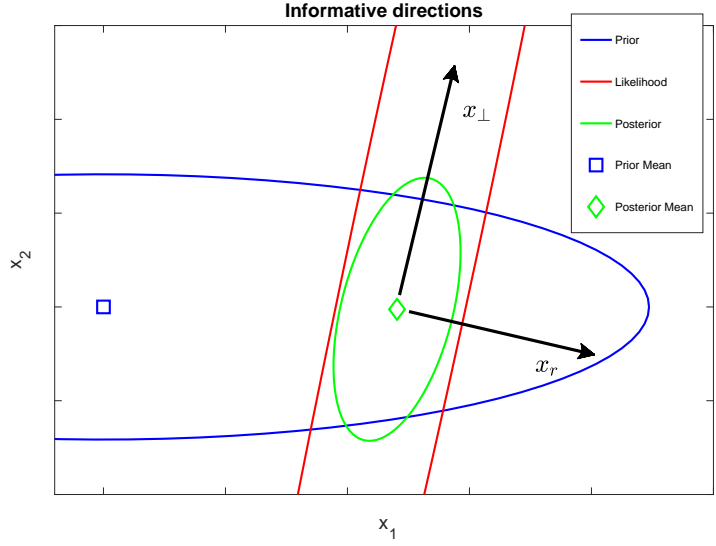}
\caption{Illustration of an informative direction $x_r$ and a non-informative direction $x_{\bot}$ using a 2-dimensional Gaussian case. Here, the likelihood has only one informative component, so the remaining direction for the posterior is obtained from the prior.}
\label{fig:0}       % Give a unique label
\end{figure}

To see the correspondence for the two approaches for the informative directions we notice that for $\widetilde{H}(x)$ it holds that
\begin{equation}
\begin{aligned}
	\Lc_{pr}^TH(x)\Lc_{pr} &=& &\Lc_{pr}^TJ(x)^T\Snoisei J(x)\Lc_{pr}^T \\
					&=& &(\Lc_{obs}^{-1}J(x)\Lc_{pr})^T(\Lc_{obs}^{-1}J(x)\Lc_{pr}) \\
					&=& &\widetilde{J}^T(x)\widetilde{J}(x).
\end{aligned}
\end{equation}
The eigenvalues $\lambda^2$ of matrix $\widetilde{H}(x)$ less than unity correspond to the singular values $\lambda$ less than unity of the scaled Jacobian $\widetilde{J}(x)$. The corresponding eigenvectors are the same as the right singular vectors $v$ of $\widetilde{J}$. The informative and non-informative directions for a simple 2-dimensional Gaussian case are illustrated in Figure~\ref{fig:0}.

Next, we use the informative directions of the measurement to reduce the dimension of the inverse problem.  Consider approximations for the posterior of the form
\begin{equation}\label{postapprox}
	\widetilde{\pi}(x|y) \propto \pi(y|\Pi_rx)\pi_{pr}(x),
\end{equation}
where $\Pi_r$ is rank $r$ projection matrix. In \cite{TC1} and \cite{TC2} it was shown that for any given $r$, there exists a unique optimal projection $\Pi_r$ that minimizes the Hellinger distance between the approximative rank $r$ posterior and the full posterior. Furthermore, using the connection to Rodgers' formalism, the optimal projection can be obtained explicitly with the following definition.

\begin{definition}[LIS]\label{lisdef}
Let $V_r\in\R^{n\times r}$ be a matrix containing the first $r$ left singular vectors of the scaled Jacobian $\widetilde{J}$. Define 
\begin{equation}
  \label{equ:phir}
\Phi_r := \Lc_{pr} V_r \text{ and } \Theta_r := \Lc_{pr}^{-T}V_r.
\end{equation}
The rank $r$ \emph{LIS projection} for the posterior approximation (\ref{postapprox}) is given by
\begin{equation}
	\Pi_r = \Phi_r\Theta_r^T.
\end{equation}
The range $\X_r$ of projection $\Pi_r: \R^n\to\X_r$ is a subspace of state space $\R^n$ spanned by the column vectors of matrix $\Phi_r$. We call the subspace $\X_r$ the \emph{likelihood-informed subspace (LIS)} for the linear inverse problem, and its complement $\R^n\setminus\X_r$ the \emph{complement subspace (CS)}.
\end{definition}

\begin{definition}\label{thedef}
The matrix of singular vectors $V = [V_r V_{\bot}]$ forms a complete orthonormal system in $\R^n$ and we can define 
\begin{equation}
  \label{equ:priperp}
\Phi_{\bot} := \Lc_{pr}V_{\bot} \text{ and } \Theta_{\bot} := \Lc_{pr}^{-T}V_{\bot} 
\end{equation}
 and the projection $\II-\Pi_r$ can be written as
\begin{equation}
	\II-\Pi_r = \Phi_{\bot}\Theta_{\bot}^T.
\end{equation}
Define the \emph{LIS-parameter} $x_r \in \R^r$ and the \emph{CS-parameter} $x_{\bot} \in \R^{n-r}$ as
\begin{equation}
	x_r := \Theta_r^Tx, \quad x_{\bot} := \Theta_{\bot}^Tx.
\end{equation}
\end{definition}
The parameter $x$ can now be naturally decomposed as
\begin{equation}
\begin{aligned}
	x=&\Pi_rx + (\II-\Pi_r)x \\
	=&\Phi_rx_r + \Phi_{\bot}x_{\bot}.
\end{aligned}
\end{equation}
Using this decomposition and properties of multivariate Gaussian distributions, we can write the prior as
\begin{equation}
	\pi_{pr}(x) = \pi_r(x_r)\pi_{\bot}(x_{\bot})
\end{equation}
and approximate the likelihood by using the $r$ informative directions,
\begin{equation}
	\pi(y | x) = \pi(y | \Phi_rx_r)\pi(y | \Phi_{\bot}x_{\bot}) \approx \pi(y | \Phi_rx_r),
\end{equation}
which leads us to the approximate posterior
\begin{equation}
	\widetilde{\pi}(x|y) = \pi(y|\Phi_rx_r)\pi_{r}(x_r)\pi_{\bot}(x_{\bot}).
\end{equation}

When the forward model is not linear, the Jacobian and Hessian matrices depend on the parameter $x$ and the criterion (\ref{raylei}) only holds point wise. To extend this local condition into a global one, we consider the expectation of the local Rayleigh quotient $\mathcal{R}(\Lc_{pr}v;x)$ over the posterior,
\begin{equation}
	\Ex[\mathcal{R}(\Lc_{pr}v;x)] = \frac{v^T \widehat{J}^T \widehat{J} v}{v^Tv}, \quad \widehat{J} = \int_{\R^n}\widetilde{J}(x)\pi(x|y)dx.
\end{equation}
The expectation is with respect to the posterior distribution, which is not available before the analysis. In practice, an estimate is obtained by Monte Carlo,
\begin{equation}\label{MCMCest}
	\widehat{J}_n = \frac{1}{n} \sum_{k=1}^n \widetilde{J}(x^{(k)}),
\end{equation}
where $x^{(k)}$ is a set of samples from some reference distribution which will be discussed later in this work. We can now use the singular value decomposition $\widehat{J}_n = W\Lambda V^T$ to find a basis for the global LIS analogously to the linear case.

The advantage of LIS dimension reduction is that it is sufficient to use MCMC to sample the low-dimensional $x_r$ from the reduced posterior $\pi(y | \Phi_rx_r)\pi_{r}(x_r)$, and form the full space approximation using the known analytic properties of the Gaussian complement prior $\pi_{\bot}(x_{\bot})$. 

\section{Results}
\label{sec:4}
\begin{figure}
\centering
\includegraphics[width=\textwidth]{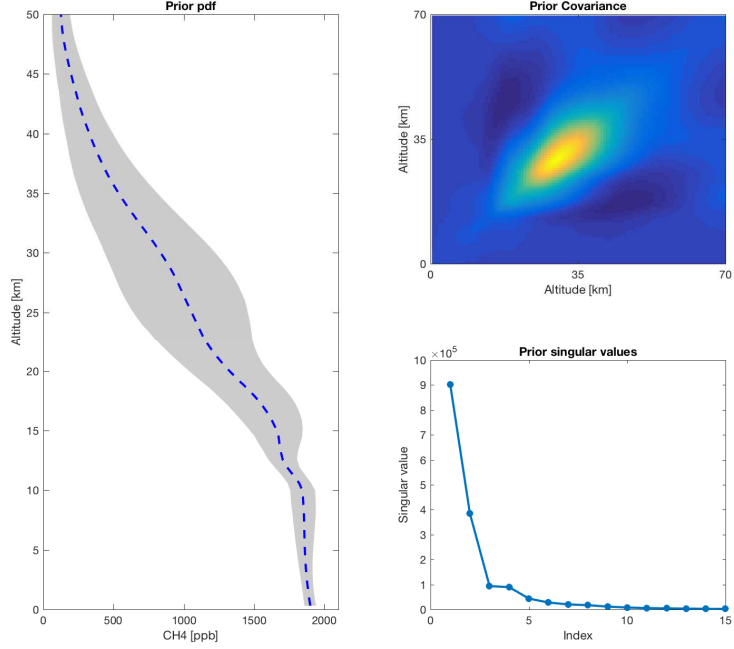}
\caption{The prior derived from an ensemble of ACE satellite measurements. Left: Full prior profile, mean with dashed line and 95\% probability limits in grey. Top right: covariance matrix derived from the measurements. Bottom right: first 20 singular values of the prior covariance matrix.}
\label{fig:1}       % Give a unique label
\end{figure}

To solve the inverse problem related to the FTIR measurement \cite{Simo}, we use adaptive MCMC \cite{HaarioEtAl2,Laine} and SWIRLAB \cite{Swirlab} toolboxes for Matlab. The results from newly implemented LIS-algorithm as well as from the previous prior reduction method are compared against a full dimensional MCMC simulation using the Hellinger distance of approximations to the full posterior.  We use a prior derived from an ensemble of atmospheric composition measurements by the ACE satellite \cite{BernathEtAl05}. The vertical prior distribution, prior covariance and prior singular values are illustrated in Figure~\ref{fig:1}. 

% which is the optimal way of choosing a prior according to \cite{Rodgers}

\begin{figure}
\centering
\includegraphics[width=\textwidth]{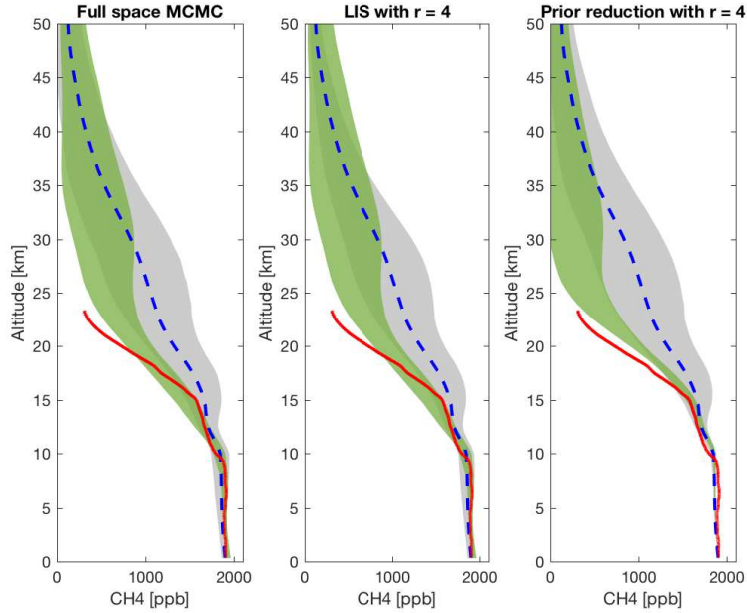}
\caption{Atmospheric $\text{CH}_4$ density profile retrieval results. Retrieved posterior in green, prior in gray, and in-situ AirCore measurement in red. The color shading indicates areas where 95\% of the profiles are. Right: MCMC with in full space. Middle: MCMC with LIS. Right: MCMC with prior reduction.}
\label{fig:2}       % Give a unique label
\end{figure}

In Figure~\ref{fig:2}, we show the results of our retrievals using full-space MCMC, compared with LIS dimension reduction and prior reduction using 4 basis vectors in each method. The retrievals are further compared against accurate in-situ measurements made using AirCore balloon soundings \cite{KarionEtAl10} which are available for the selected cases, also included in Figure~\ref{fig:2}. In this example, the Monte-Carlo estimator (\ref{MCMCest}) for $\widehat{J}_n$ in equation (\ref{MCMCest}) was computed using 1000 samples drawn from the Laplace approximation $\N(\widehat{x}, \widehat{\Sigma}_{post})$, where $\widehat{x}$ and $\widehat{\Sigma}_{post}$ are the posterior MAP and covariance, respectively, obtained using optimal estimation \cite{Rodgers}.

In order to compare the performance of MCMC methods, we define the \emph{sample speed} of a MCMC run as
\begin{definition}
The \emph{effective sample size} $N_{\rm{eff}}$ of a MCMC chain is given by
\begin{equation}
N_{\rm{eff}} = \frac{N_{\rm{M}}}{1+s\sum_{k=1}^\infty\rho_k(x)},
\end{equation}
where $N_{\rm{M}}$ is the length of the MCMC chain and $\rho_k(x)$ is lag-$k$ autocorrelation for parameter $x$ \cite{ripleystoch}.
Define the \emph{sample speed} of an MCMC chain as 
\begin{equation}
	\mathbb{V} = \frac{N_{\rm{eff}}}{t_{\rm{M}}},
\end{equation}
where $t_{\rm{M}}$ is the total computation time of the MCMC chain. 
\end{definition}

For the MCMC runs shown in Figure~\ref{fig:2}, we get as corresponding sample speeds as samples per second:
\begin{equation}
	\mathbb{V}(\text{full}) = 1.56\,\text{s}^{-1},\quad \mathbb{V}(\text{LIS}) = 19.01 \,\text{s}^{-1}, \quad \mathbb{V}(\text{PriRed}) = 19.66 \,\text{s}^{-1}.
\end{equation}

\begin{figure}
\includegraphics[width=\textwidth]{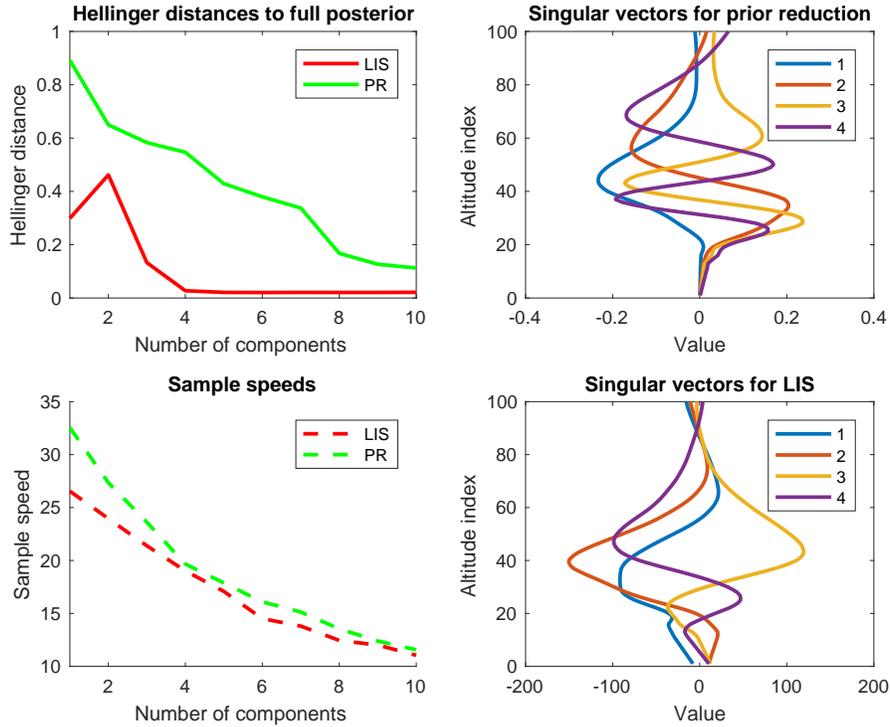}
\caption{Left: Hellinger distances to full posterior and sample speeds of corresponding MCMC runs as functions of singular vectors used in the approximation. Top right: first 4 singular vectors from prior reduction. Bottom right: first four singular vectors of $\widetilde{J}$ forming the LIS basis.}
\label{fig:3}
\end{figure}

In order to compare the approximate posteriors obtained from prior reduction and LIS-dimension reduction against the full posterior, we use the discrete Hellinger distance,
\begin{equation}
	\mathcal{H}(P,Q) = \frac{1}{\sqrt{2}}\sqrt{\sum_{i=1}^k(\sqrt{p_i}-\sqrt{q_i})^2},
\end{equation}
where $P = (p_1, \dots, p_k)$ and $Q = (q_1, \dots, q_i)$ are discrete representations of the full and approximate posterior distributions obtained from histograms of corresponding MCMC runs. The Hellinger distances of both approximations to the full posterior can be seen in Figure~\ref{fig:3} together with the corresponding sample speeds, both as a function of the number of singular vectors used. In Figure~\ref{fig:3} have also visualized the first four singular vectors used in prior reduction and LIS method for the example retrieval in Figure~\ref{fig:2}.

\section{Conclusions}
\label{sec:5}
Although both of the discussed dimension reduction methods provide roughly the same computational gains in the performance of the MCMC sampler, we see from Figure~\ref{fig:3} that while using an empirical prior, the prior reduction method requires a lot more singular vectors to achieve the same Hellinger distance from the full posterior as the LIS method, which gets really close already with 4 singular vectors. We conclude that the LIS method gives an efficient MCMC sampling algorithm to solve the inverse problem arising from the FTIR retrieval, with an additional improvement of allowing the direct usage of an empirical prior.

\begin{acknowledgement}
We thank Dr.~Rigel Kivi from FMI Arctic Research Centre, Sodankylä, Finland for the AirCore and TCCON data. We thank Dr.~Tiangang Cui from Monash University and the mathematical research institute MATRIX in Australia for organizing a workshop where a part of this research was performed. This work has been supported by Academy of Finland (projects INQUIRE, IIDA-MARI and CoE in Inverse Modelling and Imaging) and by EU's Horizon 2020 research and innovation programme (project GAIA-CLIM).
\end{acknowledgement}

%Please do not use an appendix.

%\bigskip
%
% BibTeX users please use
\bibliographystyle{spmpsci}
\bibliography{LIS_paper}

\end{document}